\documentclass[%
nofootinbib,
 amsmath,amssymb,
 aps, physrev,
 showkeys,
floatfix,
]{revtex4-2}

\usepackage{graphicx}
\usepackage{dcolumn}
\usepackage{bm}
\usepackage{hyperref}

\usepackage{xcolor}
\newcolumntype{P}{D{+}{\,\pm\,}{2}}

\begin{document}

\preprint{APS/123-QED}

\title{Student responses to a modified Force Concept Inventory: The Newtonian Mechanics Quiz}



\author{Ashutosh Kumar Pathak}
\affiliation{School of Physical Sciences, The Open University, Walton Hall, Milton Keynes, MK7 6AA, UK}

\author{Mark A. J. Parker}
\affiliation{School of Physical Sciences, The Open University, Walton Hall, Milton Keynes, MK7 6AA, UK}

\author{Andrew J. A. James}
\affiliation{School of Physical Sciences, The Open University, Walton Hall, Milton Keynes, MK7 6AA, UK}

\author{Sally E. Jordan}
\affiliation{School of Physical Sciences, The Open University, Walton Hall, Milton Keynes, MK7 6AA, UK}

\author{Jonathan Nylk}
\email[]{jonathan.nylk@open.ac.uk}
\affiliation{School of Physical Sciences, The Open University, Walton Hall, Milton Keynes, MK7 6AA, UK}

\date{\today}

\begin{abstract}

    Even in the era of modern physics, understanding and applying Newton's classical Laws of Motion are key skills for physicists and form a core component of instructional programmes.
    Students arrive in the classroom with intuitive ideas about force and motion that are often misaligned with Newtonian thinking.
    Prior research has shown that these incorrect intuitive ideas are remarkably persistent, and can stay with students throughout their education and beyond.
    Reliable instruments, that can accurately evaluate student understanding of Newtonian mechanics, are required to track student progress and evaluate the efficacy of instructional programmes.
    The Force Concept Inventory, a conceptual quiz composed of multiple-choice questions, is a well-established instrument, developed to address these challenges.
    The Force Concept Inventory, and how students respond to its questions, has been widely investigated and discussed in the literature.
    Notably, research has shown that: correct responses to questions do not always correspond to correct application of Newtonian concepts; and the limited response options given for multiple-choice questions often do not capture the full range of student thinking.
    There is therefore a burgeoning need to develop new instruments that probe students' understanding at a deeper level, by interrogating conceptual understanding from multiple perspectives and incorporating different question formats that require students to construct and consider their responses.
    We introduce the Newtonian Mechanics Quiz, a modified version of the Force Concept Inventory that incorporates follow-up questions in a variety of formats, each with different diagnostic functions.
    The Newtonian Mechanics Quiz was administered pre- and post-instruction to undergraduate physics students across six UK universities in the 2024/25 academic year.
    The student response dataset, containing 674 attempts from 450 students, is open and freely available.
    This article describes the development and validation of the Newtonian Mechanics Quiz instrument, and illustrates several novel analyses enabled by the student response dataset that can offer deep insight into students' conceptual understanding of Newton's Laws.

\end{abstract}

\maketitle

\section{Introduction}

Newton’s Laws of Motion are fundamental for understanding the physical world. As such, it is essential that students master the concepts of Newtonian mechanics during their physics studies. When students arrive in the classroom, they already have their own ideas about force and motion, derived from everyday experience involving these concepts. These intuitive ideas do not necessarily align with the concepts of Newtonian mechanics, which can lead to misconceptions and incomplete understanding when studying the subject. These misconceptions can persist even after instruction \cite{Disessa}, which prevents students from gaining mastery of Newtonian mechanics. It follows that measuring student understanding of Newtonian mechanics is an important task. This requires accurate and reliable instruments to be developed. 

The Force Concept Inventory (FCI) \cite{Hestenes} is an assessment instrument used in physics education to test for conceptual understanding of Newtonian mechanics. After minor revisions in 1995, the current version of the FCI contains 30 multiple-choice questions (MCQs), each with one correct answer, and four incorrect options -- known as distractors -- designed to align with common student misconceptions about force and motion \cite{PhysPort}. Since its introduction, the FCI has been widely used to collect data for physics education research \cite{Hake, Lasry}. Furthermore, the FCI and studies associated with it have been the subject of much discussion in the literature \cite{Huffman, Wallace, Sands, Eaton, Yasuda}.

Limitations of the MCQ format have been identified \cite{VanLehn, Nicol}. Students select from a list of pre-prepared options which may not match the student's thinking, which causes difficulties when trying to map the reasoning that students used to reach their answers \cite{Simon}. Furthermore, there exists the possibility that students could simply be guessing when answering MCQs. As a countermeasure to address these shortcomings, follow-up questions and multi-tiered assessments can be used to investigate student reasoning and understanding more rigorously \cite{Caleon, Hidayatullah}. Follow-up questions help to verify whether the answer given by a student was based on correct understanding \cite{Zhang}, while tiered assessments provide a validation of conceptual understanding by requiring answer explanations and confidence ratings \cite{Nuraeni}. 

Rebello and Zollman \cite{Rebello} investigated the functioning of the distractor options on four FCI questions. In this study, a cohort of $N = 238$ students were given free-text response (FTR) versions of four FCI items to answer. The written student responses were compared to the distractor options of the four FCI items. Rebello and Zollman found that, while there were cases where the student responses mapped to the distractor options, there were also other cases where students displayed ideas and misconceptions in their written responses that were not covered by the distractor options. These findings showed that the multiple-choice options did not account for all student reasoning and thought processes used to answer the questions. Furthermore, the Rebello and Zollman study demonstrated the method of using questions in FTR format to obtain more detailed information about student thinking.

A related study was conducted by Yasuda and Taniguchi \cite{Taniguchi}, who investigated the student reasoning used to answer two FCI items. For this study, follow-up questions were added to two FCI items, which were answered by a class of $N =111$ students. The follow-up questions were designed to probe specific knowledge that should be required to answer the question correctly. Answers to the follow-up questions revealed that students had misconceptions pertaining to applying Newton’s First and Third laws. Of note, these types of misconceptions were not covered by the corresponding distractor options. Findings from the Yasuda and Taniguchi study illustrated the benefits of using targeted follow-up questions to investigate student thought processes and understanding. 

The above discussions from the literature highlight that there exist design priorities which could be implemented in conjunction with the FCI to probe student understanding of Newtonian mechanics concepts on a deeper level.
Multiple modified versions of the FCI have demonstrated the feasibility of probing Newtonian mechanics using variants of FCI questions in different contexts \cite{McCullough}, different formats \cite{Parker2022, Parker2023}, and different representations \cite{Dancy, Nieminen}.
Using the FCI as a starting point, there is an opportunity to combine follow-up questions with questions of different formats, into one multi-modal instrument, which could test proficiency of Newtonian mechanics on a variety of different levels.

\begin{figure}[htbp]
    \includegraphics[width=0.95\textwidth]{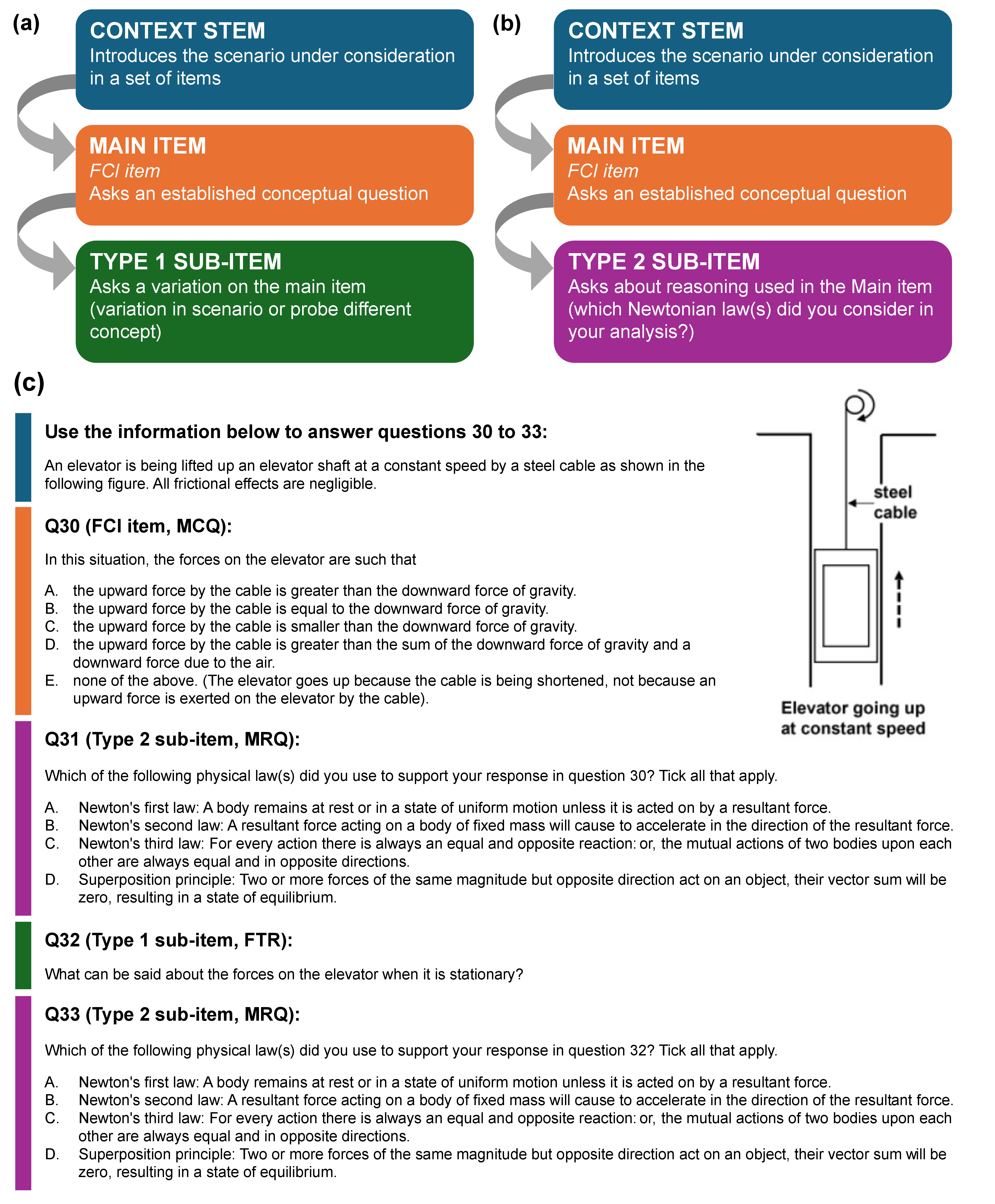}
    \caption{\label{fig:nmq_schematic} Schematic representation of the NMQ structured sub-item format.
    (a) Structure of an item group featuring an FCI item (Main item) and a Type 1 sub-item follow-up.
    (b) Structure of an item group featuring am FCI item (Main item) and a Type 2 sub-item follow-up.
    (c) Example of a question group from the NMQ (questions 30 to 33) featuring one FCI item, one Type 1 sub-item and two Type 2 sub-items.
    The Main item, Q30, is an FCI item.
    Q31 is a multiple-response Type 2 sub-item that asks about the reasoning used in answering Q30.
    Q32 is a free-text response Type 1 sub-item that asks a conceptual variation on Q30.
    Q33 is a multiple-response Type 2 sub-item that asks about the reasoning used in answering Q32.}
\end{figure}

This article introduces the Newtonian Mechanics Quiz (NMQ), a multi-modal instrument based on the FCI, which includes follow-up questions in a variety of different formats including MCQ, FTR and multiple-response question (MRQ) format.
We introduce two distinct types of follow-up questions which we term Type 1 and Type 2 sub-items.
Type 1 sub-items ask a structured follow-on question related to the previous item, that are either variations on the scenario under question or variations in question to probe a different concept related to the scenario.
Type 2 sub-items ask students to elaborate on the reasoning used to arrive at the answer to the previous item by identifying the physical principle(s) they applied in analysis of the scenario.
Items are structured into groups containing related FCI items, Type 1 sub-items and Type 2 sub-items.
A single group may contain multiple FCI items, an FCI item may be followed up by multiple related Type 1 sub-items, and a Type 2 sub-item may follow on from either an FCI item or from a Type 1 sub-item.
The typical structure of a group of related quiz items is visualized in Figure~\ref{fig:nmq_schematic}.

The purpose of the research is to use the follow-up questions of the NMQ to gain further insight into student reasoning, decision making, and thought processes when responding to Newtonian mechanics problems. Alongside this, the quantitative dataset used for the analyses presented in this article is an open and freely available dataset (see Data Availability Statement for more details) \cite{Pathak}. The dataset contains the responses given in 674 quiz attempts from 450 students who attempted the NMQ across six universities in the UK. As such, this article serves as both a descriptor of the dataset, and a demonstration of how it can be used.

\section{Development of the NMQ}\label{sec:nmq_development}

Quiz development followed the methods for designing assessment instruments outlined in the work of Crocker and Algina \cite{Crocker}, together with the concept inventory design framework outlined in \cite{Lindell}. Questions were authored for the first iteration of the quiz, with the items on the FCI (1995 version \cite{PhysPort}) used as the foundation upon which sub-questions were developed. Version 1 of the quiz contained a total of 21 questions posed in MCQ and FTR formats. The MCQs asked students to select answers from a pre-prepared set of options, while the FTR questions (Type 1 sub-items) covered similar scenarios to the MCQs, but asked students to explain their reasoning using their own words. In Version 1, there were 8 questions posed in MCQ format, and 13 questions posed in the FTR format. 

Expert review of Version 1 was conducted with a panel of 12 experienced physics educators. This led to changes in the instrument, most notably the introduction of additional follow-up questions to some of the items on the instrument to further investigate student understanding and reasoning (Type 2 sub-items). These follow-up questions were strongly recommended by the panel of experts, and the diagnostic testing aspect which underpinned the addition of these questions also aligns with previous literature \cite{Caleon}. Making these changes iterated the quiz from Version 1 to Version 2. There were 27 questions in Version 2 of the quiz, including the follow-up questions. Version 2 contained three types of questions: MCQ, FTR, and multiple response questions (MRQ). The MRQ format is similar to the MCQ format, but allows multiple options to be selected instead of just a single option.
The MRQ format was used for the newly added follow-up questions, which asked students to select which of Newton’s Laws of Motion they used in their reasoning to answer the previous question.

To check that the quiz was ready for large-scale administration, a pilot study was conducted in which the quiz was built as a Moodle quiz \cite{Moodle} and administered online via a publicly-accessible website. 43 undergraduate students attempted Version 2 of the quiz, of which 26 participants completed all 27 questions. Follow-up interviews were conducted with 9 of the participants to discuss their understanding of the questions and their responses in more detail. Interviews revealed that students viewed the FTR questions as a useful feature of the instrument. The syntax and semantics used by the students in their written responses indicated that the questions were responded to in the intended way. The qualitative data gathered from the interviews highlighted no serious issues with the quiz, and the level of the questions was found to be suitable for use with first-year college physics students. As such, the development and testing of the quiz shifted to large-scale administration of the instrument. 

By Version 2, the quiz contained a significant proportion of newly authored questions.
To aid statistical validation of the quiz, additional pre-validated questions were added to the quiz as an embedded anchor \cite{Sinharay}.
The half-length FCI (version HFCI1) \cite{Han} was used as the anchor and ensures that the quiz contains an internal reference metric for student ability that has undergone independent validation.
As this new quiz and the half-length FCI are both based on the FCI, they have good mutual conceptual alignment, meeting the requirement for a reliable embedded anchor \cite{Sinharay}.

Based on the above findings from the pilot study and interviews, and inclusion of the half-length FCI anchor questions, Version 2 of the quiz was iterated to Version 3. Version 3 of the quiz, which we refer to as the Newtonian Mechanics Quiz (NMQ), comprises 15 original FCI items in MCQ format, 11 Type 1 sub-items in a variety of formats (mainly FTR), and 7 Type 2 sub-items in MRQ format. Furthermore, 5 further original FCI items had FTR variants written and were administered to students randomly in either MCQ or FTR format.
Full details of the NMQ item listing are given in Appendix~\ref{app:nmq_design}.
Mapping the NMQ items against the six conceptual dimensions of Newtonian mechanics as identified by Hestenes, Wells and Swackhamer during the development of the FCI \cite{Hestenes} (Table ~\ref{tab:nmq_domain_coverage}) shows that sub-items are distributed throughout all six conceptual dimensions.
In summary, the NMQ is a 38-item multi-modal instrument designed to offer a multifaceted interrogation of student understanding across all conceptual domains of Newtonian mechanics.

\begin{table}[htb]
    \caption{\label{tab:nmq_domain_coverage}
    Coverage of Newtonian conceptual dimensions in the NMQ by item type.
    }
    \begin{ruledtabular}
        \begin{tabular}{lrrr}
             & \multicolumn{3}{c}{\textbf{Item count}}\\
            \textbf{Dimension} & \textbf{FCI items} & \textbf{Type 1 sub-items} & \textbf{Type 2 sub-items}\\
            \hline
            0. Kinematics & 3 & 3 & 0 \\
            1. First law & 6 & 3 & 5 \\
            2. Second law & 2 & 2 & 0 \\
            3. Third law & 3 & 1 & 2 \\
            4. Superposition principle & 2 & 1 & 3 \\
            5. Kinds of force & 9 & 3 & 0 \\
        \end{tabular}
    \end{ruledtabular}
\end{table}

\section{Administration of the NMQ}

Following finalization of the instrument, the NMQ was administered to first- and second-year physics college students at six participating UK institutions in the 2024/25 academic year. Participating institutions were recruited through professional academic networks. The NMQ was administered online, using a publicly-available Moodle website. This large-scale administration produced a response dataset for the instrument of 674 attempts from 450 students.
The distribution of student responses across the six participating institutions is shown in Table~\ref{tab:nmq_populations}.
Pre-test administration at University A was carried out in a timetabled workshop class near the start of the semester in which Newtonian mechanics is taught, with the post-test administration approximately 7 weeks later in a similar workshop session.
Pre-test and post-test attempts by the same student are linked by the same Participant ID in the dataset.
Pre-test administrations at Universities B - F were conducted as optional activities to be completed at-home near the start of the semester and had lower participation rates.
Participants had the option to disclose their gender and academic year of study at the end of the instrument.
Members of the author team ensured that the dataset was managed and stored securely as per the study's institutional ethical approval.

\begin{table}[htb]
    \caption{\label{tab:nmq_populations}
    Distribution of student responses across the six participating UK institutions in the pre- and post-test administrations of the NMQ.
    }
    \begin{ruledtabular}
        \begin{tabular}{lllr}
            \textbf{Institution} & \textbf{Institution description} & \textbf{Administration} & \multicolumn{1}{c}{\textbf{\# student attempts}}\\
            \hline
            University A & Russell group & Pre-test & 277\\
            University A & Russell group & Post-test & 245\\
            University B & Russell group & Pre-test & 63\\
            University C & Russell group & Pre-test & 59\\
            University D & Russell group & Pre-test & 12\\
            University E & 1994 group & Pre-test & 9\\
            University F & Russell group & Pre-test & 9\\
        \end{tabular}
    \end{ruledtabular}
\end{table}

All FCI items and Type 1 sub-items were marked with a binary marking scheme, Type 2 sub-items were not marked due to the nature of the items.
A score of 1 was awarded to correct answers, and a score of 0 was awarded to incorrect answers. As such, no partial credit was given. The responses to the multiple-choice questions were automatically marked by the computer, whereas the typed responses to the free-text questions were manually marked by expert human markers.

For the human marking aspect of the study, 4 physics subject experts were enlisted to mark the free-text responses. A detailed marking guide was provided for each marker. Based on this criterion, each response was marked by each marker as being either correct or incorrect. Where applicable, the overall mark used in the study was taken based on a majority ruling. For example, if a response was marked as being `correct' by 3 markers, and the same response was marked as being `incorrect' by 1 marker, then the response was adjudged to be `correct'. In the event of a tie, the final mark was decided upon after collective discussion between the markers. The objective of using this marking scheme was to base the manual marking upon a shared interpretation of the marking rules, rather than relying upon the interpretation of a single marker. 

\section{The NMQ student response dataset}

The dataset of student responses to the NMQ is provided as a single csv file.
Each row contains all the data associated with one student attempt of the NMQ.
Each row has 199 data fields covering:
\begin{itemize}
    \item Participant ID
    \item Self-reported demographic information:
    \begin{itemize}
        \item Gender
        \item Year of study
    \end{itemize}
    \item Quiz-level information:
    \begin{itemize}
        \item Host institution
        \item Type of administration (pre-test or post-test)
        \item Quiz attempt start date/time
        \item Attempt duration
        \item Total NMQ score ( /31)
        \item Half-FCI score ( /14)
    \end{itemize}
    \item For each of the 38 NMQ items:
    \begin{itemize}
        \item Item type
        \item Item format
        \item Question text
        \item Response
        \item Grade
    \end{itemize}
\end{itemize}

Note that grade information for original FCI items in MCQ format has been withheld from the dataset in an effort to maintain the test-security of the FCI.
The grading of the original FCI MCQ items can be decoded from the Response data using a copy of the FCI marking guide, available to educators and researchers on reasonable request from PhysPort \cite{PhysPort}.
The Grade fields for Type 2 sub-items are also blank, as these items were not marked.

The following sections demonstrate validity checks of the NMQ instrument and illustrate some novel analyses that are enabled by the NMQ response dataset.

\section{Validation of the NMQ}

Validity refers to verifying that an instrument measures what it was designed to. Reliability refers to the consistency of the items and instrument. The testing for NMQ validity was covered in the pilot study (Section~\ref{sec:nmq_development}), which showed that the NMQ items and overall instrument function as intended.

The testing for reliability of the NMQ was conducted by determining and evaluating quiz-level and question-level metrics from Classical Test Theory (CTT) and Item Response Theory (IRT). Throughout the CTT and IRT analyses, only quiz attempts administered pre-instruction ($N = 429$) were considered for determination of metrics unless otherwise stated. Although the NMQ contained 38 items in total, the 7 MRQ format (Type 2) follow-up questions were formative in nature, so these were not graded. As such, the NMQ was scored out of a total score of 31. There were a wide variety of scores, reflecting a cohort of students of various levels of understanding and experience.

\subsection{Classical Test Theory}

Classical Test Theory (CTT) is a statistical framework used to evaluate the functioning of assessment instruments. The statistics and methods that underpin CTT are outlined in the initial work of Crocker and Algina \cite{Crocker}, and detailed in further work \cite{Ding, Chen}. Some CTT statistics are calculated on the item level, while others are calculated on the instrument level. To evaluate the performance of the NMQ as an instrument, the CTT statistics of Difficulty, Discrimination and Kuder-Richardson Reliability were calculated. 

\subsubsection{Difficulty and Discrimination}

The Difficulty index of an item is calculated as the fraction of test-takers who answered the item correctly. It is defined as

\begin{equation}
\label{eq:difficulty}
P = \frac{N_1}{N}
\end{equation}

\noindent where $ P $ is difficulty index, $ N_1 $  is the number of correct responses, and $ N $ is the number of completed tests. A larger value for the difficulty index statistic corresponds to an easier item. The acceptable range of values for difficulty index \cite{Adegoke} are between 0.3 and 1 ($ 0.3 \leq P \leq 1 $). The range of difficulty can be split into three brackets: Easier items correspond to difficulty values above 0.7 ($ 0.7 < P $); Moderately challenging items correspond to difficulty values between 0.3 and 0.7 ($ 0.3 \leq P \leq 0.7 $); And items which are unreasonably challenging correspond to difficulty values below 0.3 ($ P < 0.3 $). In practice, having moderately challenging items on an instrument is useful, as these can discriminate between test takers of different abilities. 

The Discrimination index of an item is the ability of the item to differentiate between high scoring and low scoring test takers. To calculate the discrimination index, the group of students with the highest total test scores are compared with the group students with the lowest total test scores. In the method proposed by Kelley \cite{Kelley}, students from the highest scoring 27\% are compared with those from the lowest scoring 27\%. The discrimination index for a particular item is calculated as the difference in the proportion of students who answered the item correctly between the highest performing 27\% and the lowest performing 27\%. It is calculated using the formula: 

\begin{equation}
\label{eq:discrimination}
D = {N_H-N_L}
\end{equation}

\noindent 

\noindent where $ D $ is discrimination index, $ N_H $ is the proportion of students in the highest performing 27\% who answered the item correctly, and $ N_L $ is the proportion of students in the lowest performing 27\% who answered the item correctly. Much like the item difficulty index, the item discrimination index takes a value between 0 and 1, and the acceptable range of values for the discrimination index are 0.2 and above ($ 0.2 < D $) \cite{Ebel}.  When an item has a discrimination index of larger value, this means that the item is better at differentiating between test takers with different abilities.

\begin{figure}[htb]
    \includegraphics[width=0.85\textwidth]{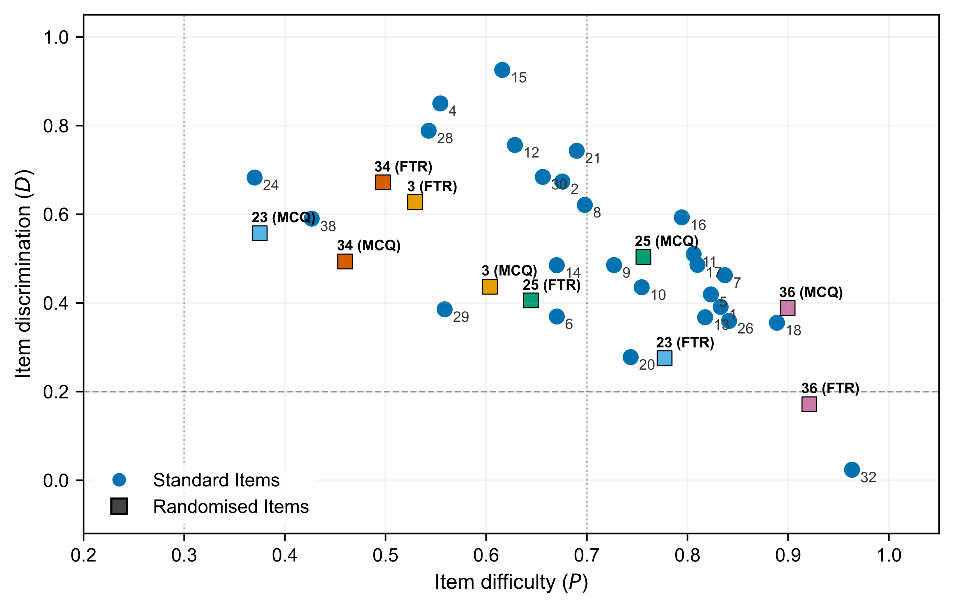}
    \caption{\label{fig:ctt_analysis} Scatter plot of item difficulty index ($P$) versus discrimination index ($D$) for all 31 dichotomously graded NMQ items. The horizontal dashed line ($D = 0.2$) separates acceptable items from those with low discriminative power. Vertical intervals filter items by difficulty tier. Unique color-coding is applied to each of the five randomized item pairs.}
\end{figure}

A scatter graph showing the difficulty and discrimination indices of the graded items from the NMQ is shown in Figure~\ref{fig:ctt_analysis}.

Figure ~\ref{fig:ctt_analysis} shows that all the items on the NMQ had difficulty values that were within the acceptable range of values ($P > 0.3$). All but two items had discrimination values that were within the acceptable range. The two items with discrimination values that fell below 0.2 were NMQ32, a Type 1 sub-item in FTR format, and the FTR version of NMQ36. In each case, almost all students who attempted the questions provided correct answers. Since everyone answers the item correctly, there is nothing  in the score to differentiate between higher and lower scoring test-takers, giving the items low discrimination capabilities. This outcome could arise because of a peculiarity in the question wording, or its context. Further work would be required to investigate these possibilities. However, since these two items were in free-text format, the written responses themselves could contain clues about students' understanding that is not captured by the item grade.

In addition to the data presented in the scatter graph in Figure CTT, the mean value of the difficulties for the graded NMQ items was calculated to be $ P_\mu = 0.69 $, while the mean value of the discrimination for these items was calculated to be $ D_\mu = 0.51 $. These values are within the acceptable range of values for difficulty and discrimination, showing that the NMQ has balanced difficulty and good discrimination capabilities as an overall instrument. 

\subsubsection{Kuder-Richardson Reliability (KR-20 Reliability)}

Reliability refers to the consistency of the items and the overall instrument. As such, students with similar knowledge and experience would be expected to get similar scores when tested with a reliable instrument. This reliability can be estimated using the Kuder-Richardson Reliability (KR-20 Reliability) statistic. The KR-20 Reliability measures the extent to which an entire test is constructed using questions that contain consistent material. It is calculated using the equation:

\begin{equation}
\label{eq:KR20}
r_{test} = \frac{K}{K-1}\Bigg(1-\frac{\sum P_i(1-P_i)}{\sigma_x^2}\Bigg)
\end{equation}

\noindent where $ K $ is the number of test items, $ P_i $ is the difficulty of the $ i^{th} $ item, and $ \sigma_x $ is the standard deviation of the total score. A value of  $ r_{test} $  of $0.8$ or above shows that the test is reliable overall \cite{Ding}.

For the full version of the NMQ, the KR-20 reliability was calculated to be $0.94$, which indicates a high level of internal consistency. When restricted to the 14 items taken from the half-length FCI, the KR-20 reliability was found to be $0.85$. Both values were above $0.8$, showing internal consistency, and demonstrating that the NMQ instrument provides a reliable and consistent measure of student understanding of Newtonian mechanics concepts. 

\subsection{Correlation between Half-length FCI score and Total NMQ score}\label{subsect:hafl-fci-corr}

\begin{figure}[b]
        \includegraphics[width=0.75\textwidth]{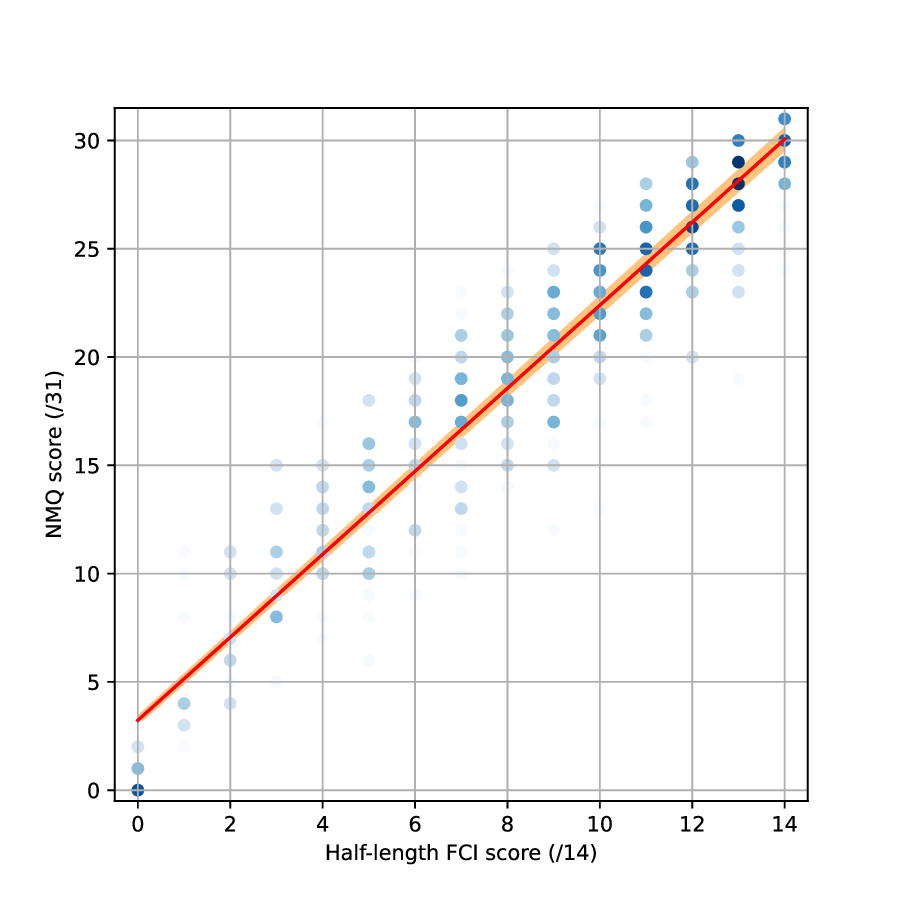}
        \caption{\label{fig:score_correlation_analysis}Graph showing the correlation between the half-FCI score and the total NMQ score. Blue circles show the correlation between half-length FCI score (out of 14) and total NMQ score (out of 31).
        Color shading indicates the density of the score distribution.
        Red line indicates a linear line of best fit with the model $y = \left( 1.92 \pm 0.02 \right) x + \left( 3.2 \pm 0.2 \right)$.}
    \end{figure}

The NMQ contains 14 questions from the half-length FCI. As a further validity check, the total scores on the NMQ (out of 31) were compared with the total scores on the half-length FCI (out of 14).
All 674 NMQ attempts were used for this analysis.
Figure~\ref{fig:score_correlation_analysis} shows the correlation between the half-length FCI score and the total NMQ score. The NMQ score is highly correlated with the score from the selection of items that make up the half-length FCI, with a Pearson linear correlation coefficient of $0.954$ ($p < 10^{-6}$). Therefore, the half-length FCI score can be used as an externally-validated internal reference measure of student ability that is independent of the new question features on the NMQ, such as the subjectivity of free-text response marking.

\subsection{Item Response Theory}

Item response theory (IRT) \cite{Wang} modeling was conducted as a complementary analysis to CTT. IRT provides further insight into the difficulty and discrimination of items across different levels of student ability.

In IRT, the probability of answering each item in a psychometric assessment is expressed as a function of the test taker's underlying ability.
In the following IRT analysis of the NMQ, we used the score achieved on the 14 half-length FCI items as an externally validated ability metric.
The half-length FCI ability metric, $\theta^{\prime}$, was then standardized through unit variance scaling to give the standardized ability metric, $\theta$:
\begin{equation}
    \label{eq:ability_scaling}
    \theta = \frac{\theta^{\prime} - \theta^{\prime}_{\mu}}{\theta^{\prime}_{\sigma}}
\end{equation}

\noindent where $\theta^{\prime}_{\mu}$ and $\theta^{\prime}_{\sigma}$ are the mean and standard deviation of the distribution of $\theta^{\prime}$.

Considering only the $429$ NMQ attempts that were administered pre-instruction, the mean ability was $\theta^{\prime}_{\mu} = 9.42$ and the standard deviation was $\theta^{\prime}_{\sigma} = 3.57$.
Therefore, the measure of ability, $\theta^{\prime}$, determined from the half-length FCI score and ranging from $0$ to $14$ corresponds to a standardized ability metric, $\theta$, ranging from $-2.59$ to $1.33$.

\begin{figure}[b]
    \includegraphics[width=0.95\textwidth]{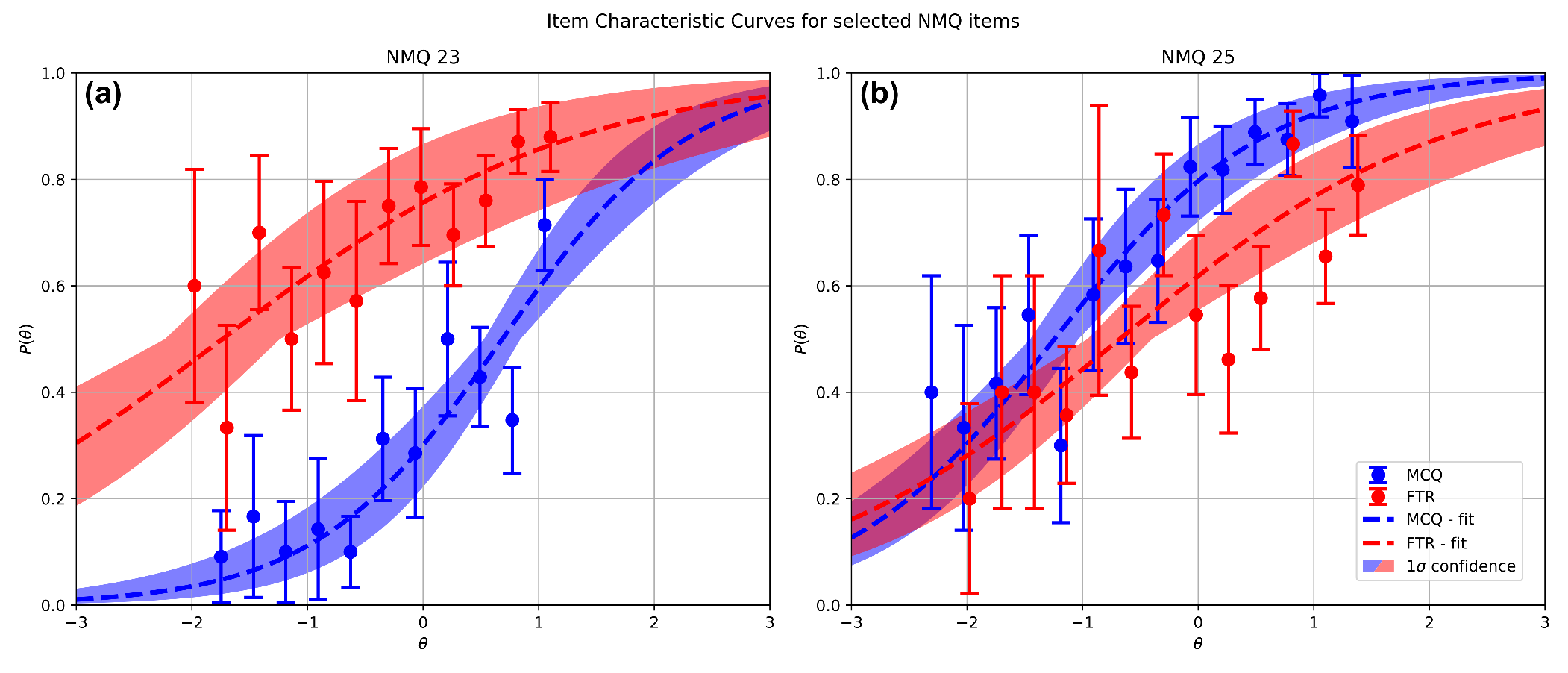}
    \caption{\label{fig:irt_analysis}
    Experimentally determined item characteristic curves (ICC) for (a) NMQ 23 and (b) NMQ 25 in MCQ (blue) and FTR (red) formats.
    Students were randomly assigned either the MCQ or FTR version of each item during administration.
    For NMQ 23, 194 students received the MCQ version, with 211 receiving the FTR version.
    For NMQ 23, 209 students received the MCQ version, with 196 receiving the FTR version.
    The probability of a student answering each item correctly as a function of standardized ability, $P(\theta)$.
    2PL IRT model fits to the data are shown by dashed lines with $1\sigma$ confidence levels in the fit parameters indicated by light color shading.
    MCQ and FTR datapoints have the same $\theta$ coordinate, but FTR data points have a small horizontal offset ($\Delta \theta = +0.05$), shifting them slightly to the right, for clarity.
    }
\end{figure}

The item characteristic curve (ICC), a graph of the probability of correctly answering an item as a function of ability, for MCQ and FTR formats of NMQ 23 and NMQ 25 are shown in Figure~\ref{fig:irt_analysis}.

The ICC can be modeled by a two parameter logistic (2PL) model.
In the 2PL IRT model, the probability of correctly answering the $i$th item is given by
\begin{equation}
    \label{eq:irt_2pl}
    P_{i} \left( \theta \right) = \frac{1}{1 - \exp \left[ -a_{i} \left( \theta - b_{i} \right) \right]}
\end{equation}

\noindent where $\theta$ represents the ability of the test taker, $a_{i}$ and $b_{i}$ are the discrimination and difficulty of item ${i}$.

The expected ICCs have a sigmoidal shape.
When the ability of a student is less than the difficulty of an item, $\theta < b$, the student has a low probability of answering the item correctly (with $P(\theta) \rightarrow 0$ for $\theta \ll b$).
When the ability of a student is greater than the difficulty of an item, $\theta > b$, the student has a high probability of answering the item correctly (with $P(\theta) \rightarrow 1$ for $\theta \gg b$).
The difficulty parameter, $b$, sets the ability level at which the transition from low to high probability occurs, and the discrimination parameter, $a$, dictates the width of the transition (with a greater value of $b$ corresponding to a sharper transition and better discrimination).
The fit parameters for NMQ 23 are, for MCQ format: $a = 1.2 \pm 0.3$, $b = 0.7 \pm 0.2$; FTR format: $a = 0.7 \pm 0.2$, $b = -1.7 \pm 0.5$.
The fit parameters for NMQ 25 are, for MCQ format: $a = 1.1 \pm 0.2$, $b = -1.2 \pm 0.2$; FTR format: $a = 0.7 \pm 0.2$, $b = -0.7 \pm 0.3$.
The two different formats of NMQ 23 show marked differences in both difficulty (horizontal translation) and discrimination (steepness/gradient).
The MCQ version of NMQ 23 is substantially more difficult and more discriminating than its FTR counterpart.
In contrast, the two different formats of NMQ 25 are quite similar in difficulty, with the FTR version being slightly more difficult than the MCQ version.
Although IRT and CTT discrimination and difficulty are not directly comparable, CTT and IRT analyses of NMQ 23 and NMQ 25 yielded similar conclusions.
While CTT offers useful insights into quiz performance, the IRT analysis, stratified by student ability, can uncover subtle nuances that are overlooked when all response data is aggregated.
Table~\ref{tab:nmq_irt} in Appendix~\ref{app:irt} lists the difficulty and discrimination for all NMQ items as determined from ICC fitting.
All NMQ items have ICCs that follow the expected 2PL model shape and yielded fitted difficulty and discrimination parameters in acceptable ranges.
Comparison of ICC fit parameters for the five NMQ items administered in randomized format indicates that item format does not produce a consistent systematic shift in item difficulty and discrimination.
Some FTR items are more difficult than their MCQ counterparts while others are less difficult, and some are more discriminating while others are less discriminating.

\section{Analyses enabled by the NMQ}\label{sec:analyses}

In this section, we outline some novel analyses that are enabled by the NMQ student response dataset:
\begin{itemize}
    \item A comparison between student response patterns to multiple-choice and free-text question formats, through analysis of an FCI item administered to students in a randomized format.
    \item An investigation of how students' free-text responses change with instruction, through analysis of pre- and post-test administrations of an NMQ item in FTR format.
    \item A deeper investigation into students' Newtonian reasoning through analysis of a question--sub-question pair, an FCI item and the relevant Newtonian law(s) selected by students in the corresponding Type 2 sub-item.
\end{itemize}

In this section, only quiz attempts administered at University A were considered to allow comparison of pre-test ($N = 277$) and post-test ($N = 245$) responses.

\subsection{Influence of question format on student response patterns}

One way to investigate student responses to NMQ items in FTR format is to compare the written responses to these items with the options presented in the MCQ versions of the same items. Findings in the literature have shown that FCI distractors -- incorrect options with plausible but non-Newtonian reasoning --  might not function in the intended manner. There are cases where certain distractor options are rarely selected at all, even if students have misconceptions that are linked to these distractors \cite{Rebello}. Furthermore, data gathered during other studies \cite{Parker} showed that students provide a variety of answers, both correct and incorrect, when the FCI questions were posed in free-text format. We conducted a mapping of incorrect free-text responses to an FTR version of an FCI item onto the distractors given in the MCQ version of the item.

We present findings for NMQ 03 here.
NMQ 03 was one of five FCI items administered to students in either MCQ or FTR formats.
The random selection occurred when the student loaded each individual quiz item and the probability of selecting each format was equal.
As such, students who were assigned an item in one format for the pre-test were not necessarily assigned the item in the same format in the post-test.
For this analysis, we further restrict our sample to students who were assigned NMQ 03 in the same format in both the pre-test and in the post-test.
We therefore analyze a sample of $93$ students, $50$ who received NMQ 03 in MCQ format and $43$ who received NMQ 03 in FTR format.

\begin{figure}[t]
    \includegraphics[width=0.75\textwidth]{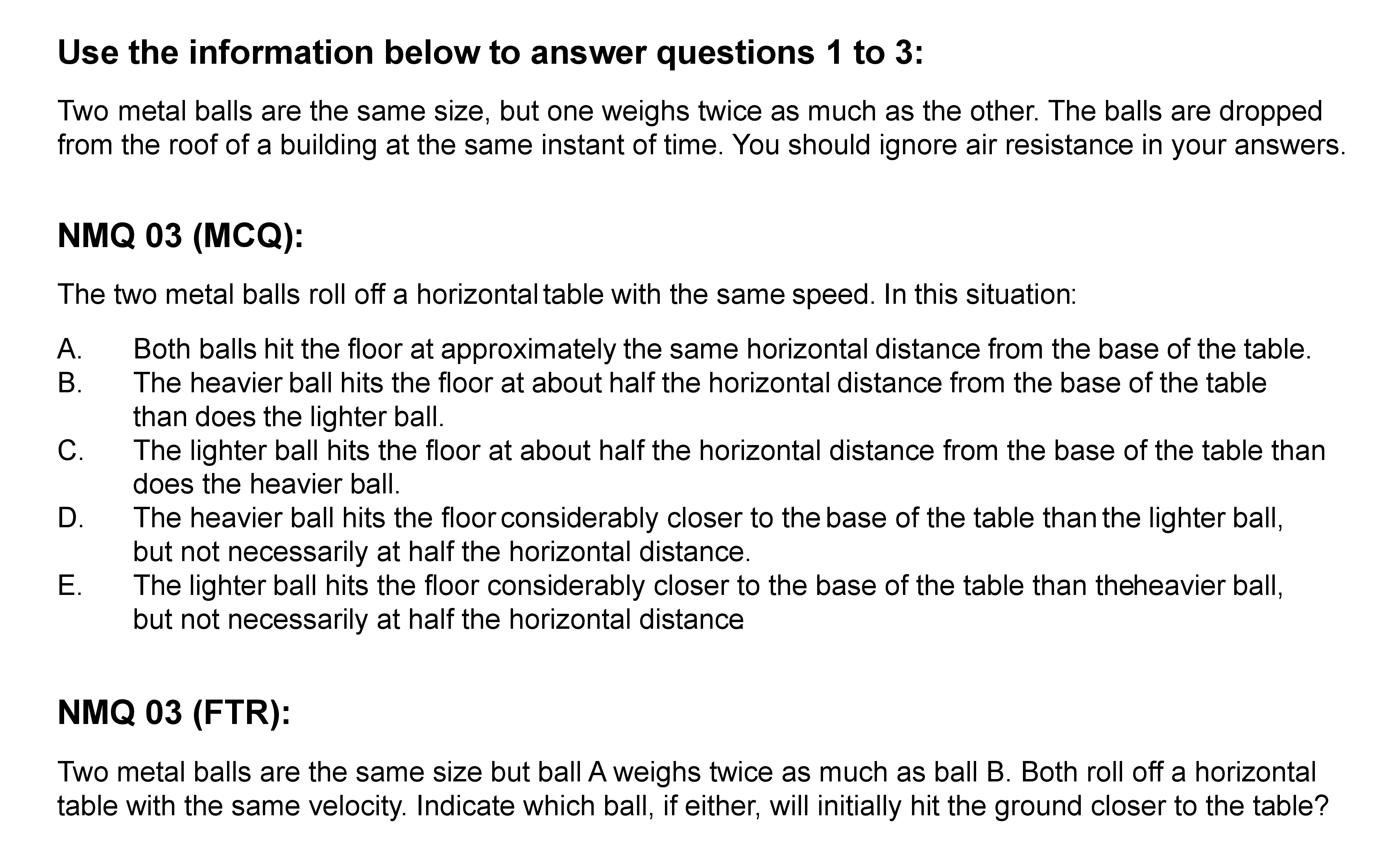}
    \caption{\label{fig:NMQ3}Context stem used for NMQ items 1 -- 3 (top), MCQ version of NMQ 03 (middle), and FTR version of NMQ 03 (bottom).}
\end{figure}

NMQ 03 asks students to identify which of two balls will land closer to a table, after both balls have rolled off the table in the horizontal direction. The question corresponds to item 2 of the original FCI.
Figure~\ref{fig:NMQ3} shows the wording of the MCQ version of NMQ 03, and the options available for students to select from, and the wording of the FTR version for comparison.

\begin{figure}[b]
        \includegraphics[width=0.95\textwidth]{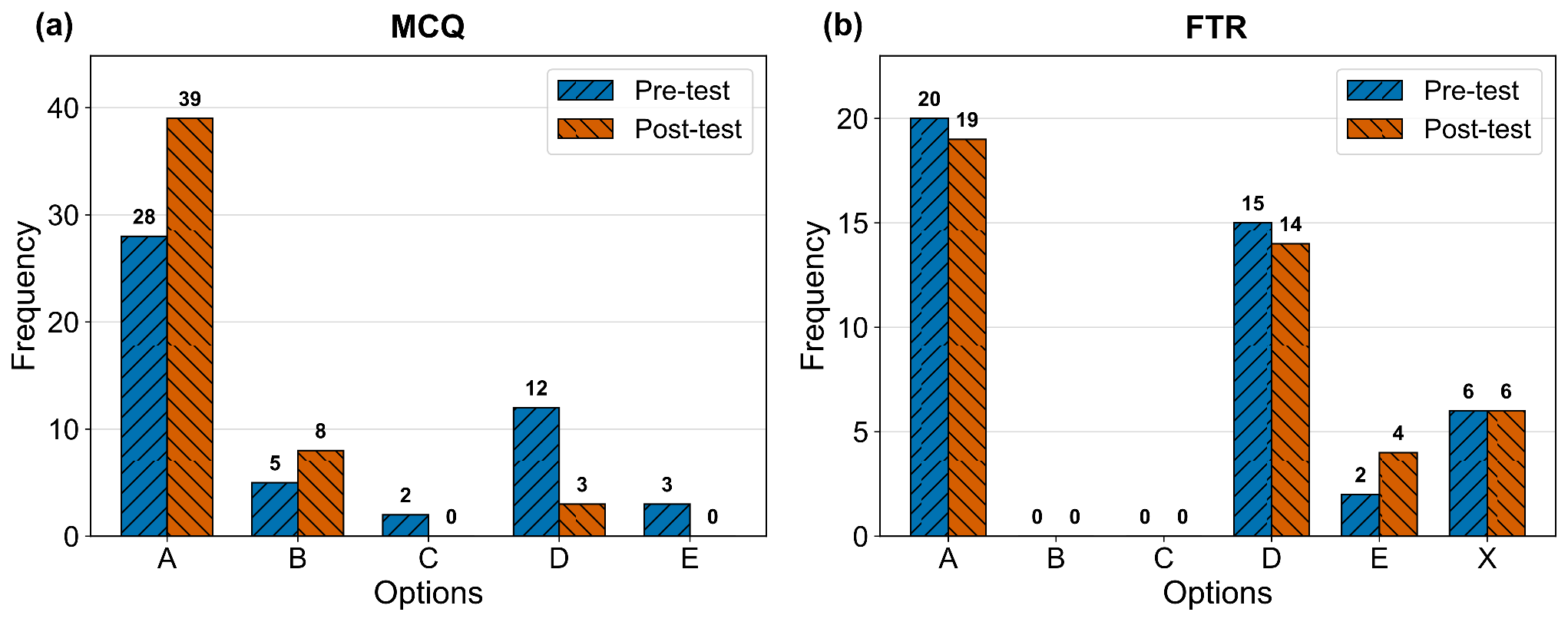}
        \caption{\label{fig:distractor_analysis} Comparison of pre-test and post-test performance for multiple-choice (a) and free-text (b) versions of NMQ 03.}
    \end{figure}

For the distractor analysis, each incorrect free-text response was mapped to a corresponding distractor, based on its contents. For cases where a feasible match was not possible, a new category, dubbed `Category X', was created. Answers placed in Category X contained novel ideas that were not covered in the distractor options. The results from these findings for NMQ 03 are shown in Figure~\ref{fig:distractor_analysis}.

For the MCQ version of NMQ 03, students mostly selected option A, which is the correct answer. Of the incorrect responses, the majority of students selected distractors B or D. The responses given to the FTR version of the question follow a similar pattern. In this case, students mostly either wrote a correct answer which could be mapped onto option A, or wrote an incorrect answer which could be mapped to distractor option D. Of note, none of the free-text responses mapped to the distractor options B or C. 

Category X, which contained more than $10\%$ of all free-text responses, contained incorrect answers that mentioned time instead of distance. NMQ 03 does not ask about time, but some students provide answers as if the question did ask about time. The distractors in the MCQ version of this item do not account for incorrect answers of this type. As such, this particular kind of incorrect answer is only revealed through the free-text responses. It is worth noting that the two items before NMQ 03, namely NMQ 01 and NMQ 02, are based upon a scenario involving free-fall and ask about the time of flight for different trajectories, and this may have affected subsequent responses to NMQ 03. This could be an example of an order effect, which is a phenomenon which occurs when the sequence of test items influences student responses as they progress through the instrument \cite{Gray, Weinstein}. Further work is required to explore this possibility.

The findings from the distractor mapping analysis pertaining to NMQ 03 can be summarized as follows. By comparing the responses from the FTR format with the corresponding MCQ format options, the analysis examined whether the distractor items mapped to ideas and misconceptions that the students had. Some distractor options were almost never selected, despite these being based upon related misconceptions that students often have. In contrast, the student responses to the FTR format showed a variety of ideas and reasoning, including an issue with understanding the question which was not captured in the distractor options. This indicates that the MCQ format can fail to present the full range of ideas students have about a item, which have been demonstrated in student free-text responses.  

\subsection{Evolution of language patterns with instruction: Evidence from free-text responses}

Analysis of the written responses to the FTR format items on the NMQ provides an opportunity to examine how students construct their written responses before and after instruction. With this idea as a basis, examining item response length and comparing this to student performance between pre-test and post-test is one way to probe changes in student learning and understanding. Word count analysis can provide an indication of whether students expand, refine, or condense their written responses post-instruction. As such, the rationale for this part of NMQ analysis was to use word count analysis to investigate how student answer length relates to performance level. This provides a method to examine how students develop their answers from pre-instruction to post-instruction. 

Here we present findings for NMQ 34, an FCI item posed in randomized format.
Again, we restrict the analysis to the sample of $47$ students who received NMQ 34 in FTR format in both pre- and post-test administrations.

\begin{figure}[b]
        \includegraphics[width=0.75\textwidth]{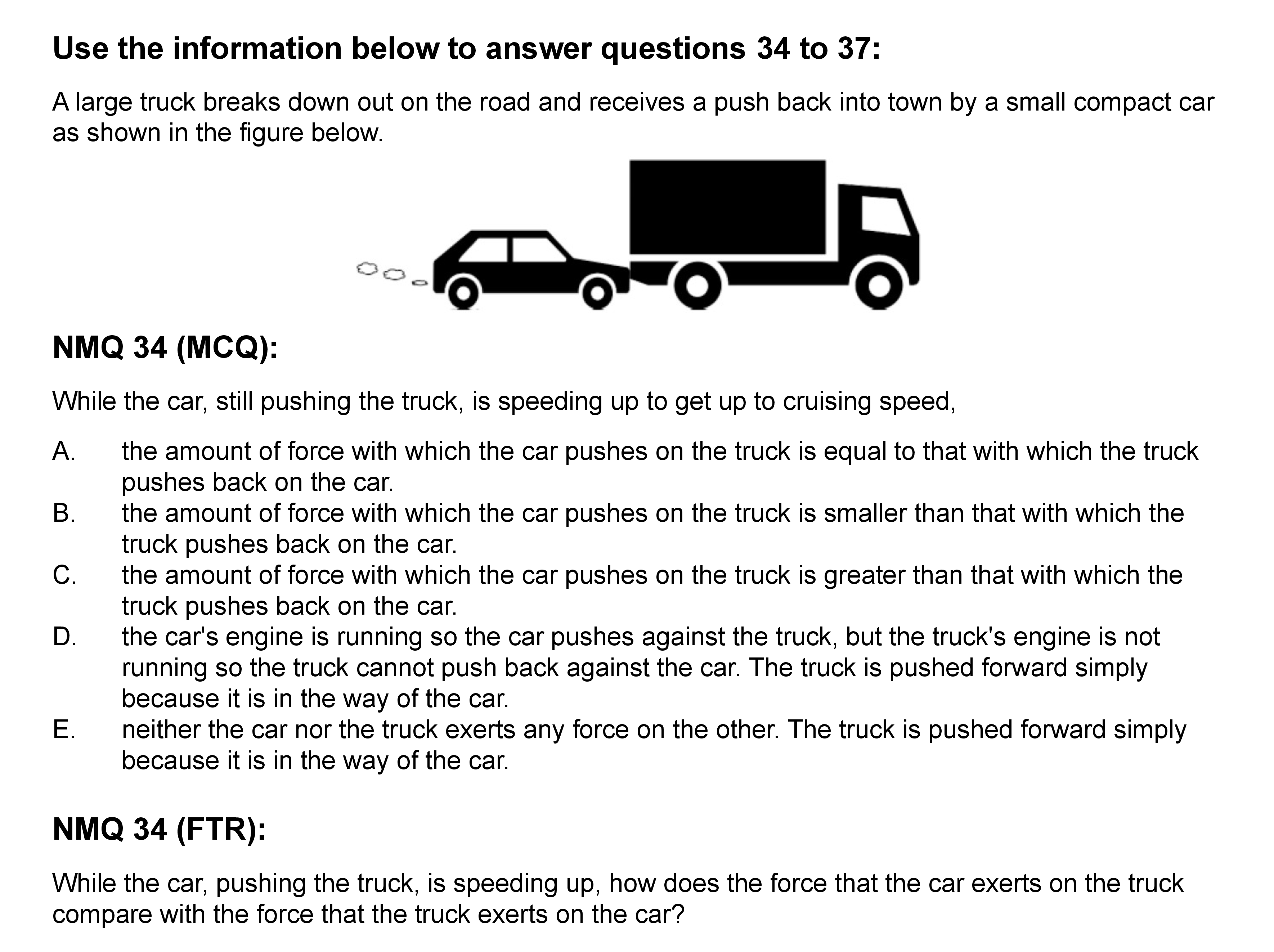}
        \caption{\label{fig:NMQ34} Context stem used for NMQ items 34 -- 37 (top), MCQ version of NMQ 34 (middle), and FTR version of NMQ 34 (bottom).}
    \end{figure}

NMQ 34 asks students to compare the force that an accelerating car pushing a truck exerts on the truck, to the force that the truck exerts back on the car. The question is adapted from Q15 of the original FCI. Both versions of the question are shown in Figure~\ref{fig:NMQ34}.

The number of words that the students used in each of their pre-test and post-test answers was counted. It was also recorded whether the student answered the question correctly or incorrectly in the pre- and the post-tests and students were categorized based on their progression.
Students who answered the item correctly on both the pre-test and the post-test were classed in the `Maintaining Mastery' category; Students who answered the item incorrectly on the pre-test, then answered correctly on the post-test were classed in the `Successful Learning' category; Students who answered incorrectly on both the pre-test and the post-test were classed in the `Persistent Difficulties' category. No student answered the item correctly on the pre-test, then incorrectly on the post-test.
The results from these findings are shown below, in Figure~\ref{fig:word_count_analysis}.

    \begin{figure}[t]
        \includegraphics[width=0.75\textwidth]{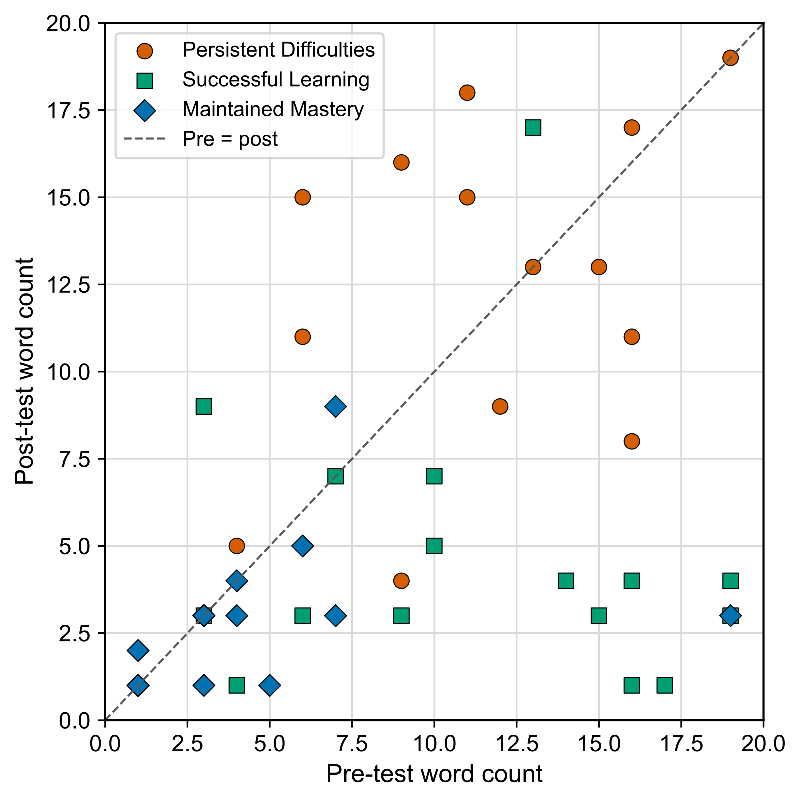}
        \caption{\label{fig:word_count_analysis} Comparison of response word count for NMQ 34 across pre- and post-test administration across for performance categories.}
    \end{figure}

Students in the `Successful Learning' category reduced their mean word count from 10.8 words in the pre-test, to 4.9 words in the post-test. The students in this category went from answering incorrectly with more words in the pre-test, to answering correctly and concisely in the post-test. This implies that improved understanding for this item was correlated with shorter, focused answers. 

Students in the `Maintaining Mastery' category also produced slightly shorter answers for this item in the post-test, where the mean answer length decreased from 4.7 words in the pre-test, down to 2.9 words in the post-test. This suggests that students can still become more proficient in expressing their understanding in a concise manner, even if they had mastery of the topic before instruction. Of note, responses which were between one and six words in length were marked as correct $90\%$ of the time in the post-test. This supports the idea that students are more capable of expressing their reasoning in short, concise statements after instruction.

On the other hand, students in the `Persistent Difficulties' category slightly increased their mean answer length from 11.6 words in the pre-test to 12.4 words in the post-test. These answers were longer than the corresponding answers given by students in the other learning categories. The consistent pattern in these incorrect answers might arise owing to the students having an incomplete understanding of the topic, leading to longer answers which contain irrelevant or incorrect information. 
 
This result demonstrates the potential for the content of students' free-text responses to provide insight into their understanding that would be obscured by only considering the correct/incorrect classification of the response.
When only considering the marking classification of a response, students in the `Maintained mastery' category would show no change of understanding in relation to NMQ 34 over the course of instruction.
However, analysis of free-text word count, and further analysis of response content and language, can reveal deeper changes in student thinking and how they construct their responses.

Taking the distractor mapping analysis findings together with the word count analysis findings, the analysis of free-text responses offered insight into how students thought about answering the questions. These ideas were sometimes concise or verbose; detailed or vague, based on the level of understanding that the students had of the topics being tested. In addition, the free-text responses revealed ideas and patterns of reasoning that were not always captured within the distractor options. Collecting this together, these findings suggest that the FTR format is useful for learning about student thought processes, as it uncovers different kinds of student reasoning and understanding.

\subsection{Paired question analysis of student response and reasoning}

    The Type 2 sub-items of the NMQ ask students to expand on their reasoning for the previous item by stating which Newtonian law(s) were used to arrive at their response.
    The wording used for Type 2 MRQ sub-items is shown in Figure~\ref{fig:nmq_schematic}(c) which shows NMQ 31, the Type 2 sub-item of NMQ 30, and NMQ 33, the Type 2 sub-item of NMQ 32.
    Comparing the reasoning given in the Type 2 sub-item and relating it to the response given in the corresponding `Main' item can provide insight into whether a student is correctly applying the relevant physical principle or incorrectly applying an irrelevant physical principle.

    We present findings relating to NMQ 36, an FCI item, and NMQ 37, its corresponding Type 2 sub-item.
    The majority of students answer NMQ 36 correctly (CTT difficulty index: $P_{36} > 0.9$, see Figure~\ref{fig:ctt_analysis}) in both MCQ and FTR formats.
    Due to the high proportion of students who answer NMQ 36 correctly, comparison of pre-test and post-test results show little change, making NMQ 36 ineffective for measuring a learning gain with instruction.
    Both formats of NMQ 36 were aggregated for analysis, and we only consider students who correctly answered the item.
    For these students, we investigated their pre-test ($N = 243$) and post-test ($N = 219$) responses to NMQ 37 to better understand their thought processes in relation to answering NMQ 36.

\begin{figure}[t]
        \includegraphics[width=0.95\textwidth]{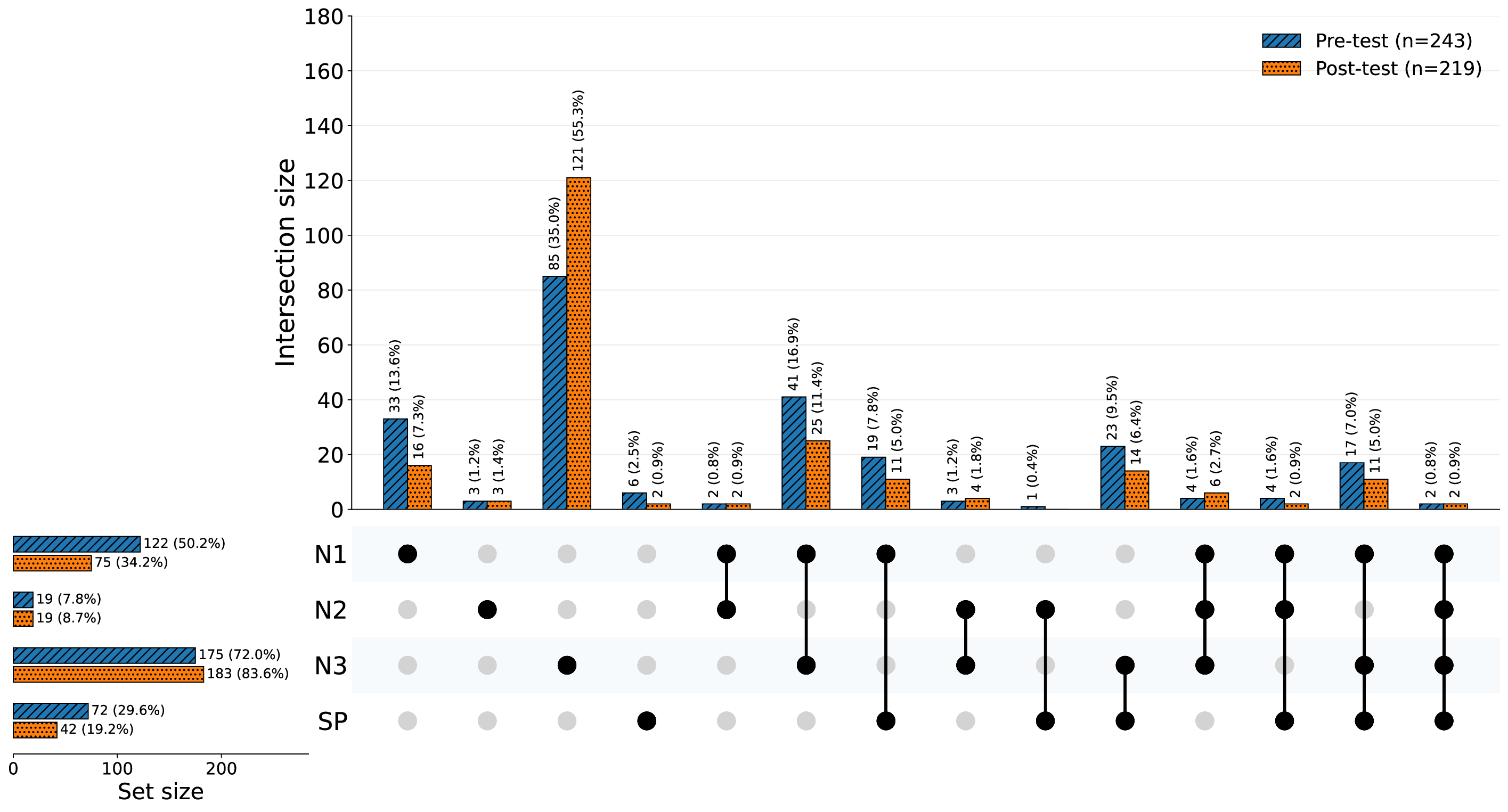}
        \caption{\label{fig:lawselection_analysis} UpSet plot showing responses to NMQ 37 (Type 2 sub-item) indicating reasoning, via Newtonian law selection, used in answering NMQ 36 (FCI item in randomized MCQ or FTR format) in pre- (blue, diagonal shading) and post-test (orange, dotted shading) NMQ administrations.
        Responses are only shown for participants that correctly answered NMQ 36.
        N1: Newton's first law; N2: Newton's second law; N3: Newton's third law; SP: Superposition principle.}
    \end{figure}

    Figure~\ref{fig:lawselection_analysis} shows the results as UpSet plots \cite{Lex}.
    The bottom right section of the plot (the combination matrix) shows the different combinations of laws given in student responses, represented by a column of shaded circles and connecting vertical lines.
    The horizontal bars to the left of the combination matrix represent the total number of students who selected each law (irrespective of whether they had selected the law alone or in combination with others) while vertical bars above the combination matrix show the number who gave each specific combination.
    For example, Figure~\ref{fig:lawselection_analysis} shows that, of the students who answered NMQ 36 correctly, 19 students reported use of Newton's second law (N2) in their reasoning in the pre-test (second row down, blue horizontal bar with diagonal shading), while only 6 students reported using the specific combination of Newton's first, second and third laws (N1, N2 and N3) in their reasoning in the post-test (fourth column from the right, orange vertical bar with dotted shading).

    NMQ 36 is similar to NMQ 34 and is adapted from Q16 of the original FCI.
    NMQ 36 differs from NMQ 34 (see Figure~\ref{fig:NMQ34}) in that the car pushing the truck is now moving at a constant velocity, but students are still asked to compare the forces that the car and truck exert on each other.
    The question is a classic example of action-reaction pairs as described by Newton's third law.
    $35.0\%$ and $55.3\%$ of students, in the pre-test and post-test respectively, make exclusive use of Newton's third law to answer NMQ 36, and $72.0\%$ and $83.6\%$ of students, in the pre-test and post-test respectively, make use of Newton's third law in combination with other reasoning.
    This shows a positive trend with instruction of students' ability to correctly identify the relevance of Newton's third law to this scenario and apply it.

    In the pre-test, there is a large population of students who correctly answer NMQ 36 but use Newton's first law in their reasoning ($13.6\%$ sole use, $50.2\%$ combined use).
    This indicates that these students may be confusing equal-and-opposite reaction pairs between two objects with two balanced forces acting on a single object.
    Following instruction, the reported use of Newton's first law fell substantially ($7.3\%$ sole use, $34.2\%$ combined use), but still remains part of the reasoning for over a third of students who correctly answer NMQ 36.
    This point is further investigated, drawing on additional evidence, in an accompanying article \cite{PathakN3}. 

\section{Conclusion}

This article introduced the NMQ, a multi-modal instrument designed to test for conceptual understanding of Newtonian mechanics. The NMQ was based upon a modified version of the FCI, and included questions from the Half-length FCI \cite{Hestenes, Han}. With this as the starting point, the NMQ was designed to include questions in different formats, and follow-up questions to the main items. Alongside the instrument, the dataset comprising of $N = 647$ students attempts of the NMQ was also introduced. The NMQ was validated through a combination of a pilot study, CTT analysis, and IRT analysis. The pilot study showed that the items on the NMQ and the overall instrument functioned as intended. Results from the CTT analysis showed that the NMQ is a reliable instrument, both at the item level and overall. Findings from the IRT analysis provided further insights into the behavior of NMQ items.
Both CTT and IRT analyses were able to detect the same shifts in item difficulty and discrimination when items were administered in different formats.
Item format does not appear to induce a systematic shift in difficulty or discrimination, with some FTR items being more difficult and some less difficult, and some more and some less discriminating, than their MCQ counterparts.
Grading of free-text responses is inherently subjective, and future research should seek to address the impact of marking guides and marker judgment on the interpretation of free-text responses and on the perceived difficulty and discrimination of FTR items.

A subset of the dataset drawn by matching pre-test ($ N = 277 $) and post-test ($ N = 245 $) student attempts were used to conduct various types of analysis. The Distractor Analysis presented a comparison of FTR and MCQ versions of an FCI item (NMQ 03). This analysis found that students used a variety of different ideas when answering the FTR version of the item, including ideas which were not captured in the distractor options of the MCQ version. The Word Count Analysis focused on the FTR version of an FCI item (NMQ 34). For this item, students went from frequently providing answers which were protracted and incorrect in the pre-test, to frequently providing answers that were concise and correct in the post-test. In addition, some responses implied that the order of the questions might have affected the way that students expressed their answers. Taken together, these findings highlight that free-text responses offer extra dimensions, such as word count or word content, to measure student understanding beyond the dichotomous correct/incorrect classification.

The Law Selection Analysis investigated the responses given to a main NMQ item (NMQ 36) in relation to its corresponding Type 2 MRQ sub-item (NMQ 37). This analysis revealed cases where students answered the main item correctly, but responses to the sub-item indicated that these answers were reached using reasoning that was not correct. These findings illustrate that sub-items allow for deeper investigation into the diagnostics of student reasoning and responses. 

Combining the Law Selection Analysis findings with the above findings from analysis of students' free-text responses, the key outcomes from the NMQ analysis can be summarized as follows. Three types of study were presented in this article: Comparing FTR responses to MCQ distractor choices; Analyzing specific patterns in word and answer length; And investigating student answers and follow-up explanations. Although different, each of these investigations has shown the capability to provide physics educators with useful information about where current forms of assessment function well, and where they can be improved upon.

Based on advice from the literature, PhysPort has developed an evaluation framework for concept inventories \cite{Madsen, PhysPort}.
The development of the NMQ was motivated by open questions arising from research into student responses to the FCI and related topics, has involved interviews with students and subject experts, administration at multiple UK institutions, and is supported by statistical analysis of item functioning.
While administration has involved six UK universities, we acknowledge that post-test data, on which the analyses presented in Section~\ref{sec:analyses} are based, was only collected from one institution.
Wider post-test administration of the NMQ across the UK, and use of the NMQ internationally, would allow further validation of the instrument and strengthen the findings presented here.
We warmly invite the Physics Education Research community to engage in the wider rollout of the NMQ, and further analysis of what student responses to the NMQ can uncover about student understanding of Newtonian mechanics.

We expect the NMQ student response dataset, and the analyses it enables, will be useful for those working in the fields of Physics Education Research, Automatic Short Answer Grading \cite{lade-etal-2026-domain} -- as well as Natural Language Processing more generally -- and Assessment Design. We anticipate translation of the NMQ instrument design and findings from the student response dataset will lead to robust assessment instruments across Science, Technology, Engineering and Mathematics disciplines. 

\section*{Acknowledgments}

The authors thank the students who participated in pilot testing and interviews as part of development of the Newtonian Mechanics Quiz.
We also thank the many colleagues who have given willingly of their time to advise, formally and informally, on the development of the Newtonian Mechanics Quiz.

\section*{Data availability statement}

The NMQ student response dataset described in this work is freely available via the UK Open University's Open Research Data Online repository \cite{Pathak}.

Grade information for original FCI items in MCQ format has been withheld
from the dataset in an effort to maintain the test-security of the FCI.
The grading of the original FCI MCQ items can be decoded from the Response data using a copy of the FCI marking guide, available to educators and researchers on reasonable request from PhysPort \cite{PhysPort}.

\section*{Ethics statement}

The study was approved by the UK Open University’s Human Ethics Research Committee, reference 2024-0387-3.

\bibliography{apssamp}

\appendix

\newpage
\section{Additional description of NMQ instrument}\label{app:nmq_design}

Appendix~\ref{app:nmq_design} contains additional additional descriptions of each NMQ item.

For each NMQ item, Table~\ref{tab:nmq_description}
states whether the item was taken from the FCI or if it was a newly authored Type 1 or Type 2 sub-item.
The Table also states the format in which each item was served to students, and which conceptual dimension(s) of Newtonian mechanics each item relates to.
Where an item is found in the FCI, the original FCI item number is given, and whether the item is found in the half-length FCI.

\begin{table}[b]
    \caption{\label{tab:nmq_description}
    Summary of NMQ item composition, detailing the type and format for all 38 NMQ items.
    For items taken from the FCI, the original FCI item number is stated to facilitate comparison with other studies.
    FCI items contained within the NMQ are additionally labeled to indicate their inclusion in the half-length FCI.
    Items are also labeled with the conceptual dimensions on Newtonian mechanics they relate to.
    MCQ: multiple-choice question; MRQ: multiple-response question; FTR: free-text response question; Y/N: yes/no question.
    Conceptual dimensions: 0.~Kinematics; 1.~First law; 2.~Second law; 3.~Third law; 4.~Superposition principle; 5.~Kinds of force.
    }
    \begin{ruledtabular}
        \begin{tabular}{llllll}
            \textbf{Item \#} & \textbf{Type} & \textbf{Format} & \textbf{FCI \#} & \textbf{In Half-FCI?} & \textbf{Dimension}\\
            \hline
            NMQ 01 & FCI item & MCQ & 1 & N & 5 \\
            NMQ 02 & Type 1 sub-item & FTR & - & - & 5 \\
            NMQ 03 & FCI item & MCQ/FTR (randomized) & 2 & Y & 5\\
            NMQ 04 & FCI item & MCQ & 5 & Y & 5 \\
            NMQ 05 & Type 1 sub-item & FTR & - & - & 0 \\
            NMQ 06 & Type 1 sub-item & FTR & - & - & 2 \\
            NMQ 07 & FCI item & MCQ & 6 & Y & 1 \\
            NMQ 08 & Type 1 sub-item & FTR & - & - & 5 \\
            NMQ 09 & FCI item & MCQ & 8 & Y & 1, 2 \\
            NMQ 10 & FCI item & MCQ & 9 & Y & 0 \\
            NMQ 11 & FCI item & MCQ & 10 & Y & 1 \\
            NMQ 12 & FCI item & MCQ & 11 & Y & 4, 5 \\
            NMQ 13 & FCI item & MCQ & 19 & Y & 0 \\
            NMQ 14 & FCI item & MCQ & 12 & Y & 5 \\
            NMQ 15 & FCI item & MCQ & 13 & Y & 5 \\
            NMQ 16 & FCI item & MCQ & 28 & Y & 3 \\
            NMQ 17 & Type 1 sub-item & FTR & - & - & 0 \\
            NMQ 18 & FCI item & MCQ & 7 & N & 1 \\
            NMQ 19 & Type 2 sub-item & MRQ & - & - & 1 \\
            NMQ 20 & Type 1 sub-item & Y/N & - & - & 1 \\
            NMQ 21 & FCI item & MCQ & 25 & N & 1, 4 \\

            NMQ 22 & Type 2 sub-item & MRQ & - & - & 1, 4 \\
            NMQ 23 & FCI item & MCQ/FTR (randomized) & 26 & Y & 0 \\
            NMQ 24 & Type 1 sub-item & FTR & - & - & 2 \\
            NMQ 25 & FCI item & MCQ/FTR (randomized) & 27 & N & 5 \\
            NMQ 26 & Type 1 sub-item & FTR & - & - & 1 \\
            NMQ 27 & Type 2 sub-item & MRQ & - & - & 1 \\
            NMQ 28 & FCI item & MCQ & 18 & N & 5 \\
            NMQ 29 & Type 1 sub-item & FTR & - & - & 0 \\
            NMQ 30 & FCI item & MCQ & 17 & Y & 1, 4, 5 \\
            NMQ 31 & Type 2 sub-item & MRQ & - & - & 1, 4  \\
            NMQ 32 & Type 1 sub-item & FTR & - & - & 1, 4, 5 \\
            NMQ 33 & Type 2 sub-item & MRQ & - & - & 1, 4 \\
            NMQ 34 & FCI item & MCQ/FTR (randomized) & 15 & Y & 3 \\
            NMQ 35 & Type 2 sub-item & MRQ & - & - & 3 \\
            NMQ 36 & FCI item & MCQ/FTR (randomized) & 16 & N & 3 \\
            NMQ 37 & Type 2 sub-item & MRQ & - & - & 3 \\
            NMQ 38 & Type 1 sub-item & FTR & - & - & 3 \\
        \end{tabular}
    \end{ruledtabular}
\end{table}


\newpage
\section{Full IRT results - difficulty and discrimination parameters}\label{app:irt}

Appendix~\ref{app:irt} contains further details of difficulty and discrimination parameters obtained from fitting a 2PL-IRT model (equation~\eqref{eq:irt_2pl}) to the ICC for each of the 31 dichotomously marked NMQ items.

All NMQ items had ICCs that followed the expected sigmoid shape and fitting was completed successfully for all items. Table~\ref{tab:nmq_irt}
lists the discrimination, $a$, and difficulty, $b$, for each item.
Where an item is found in the FCI, a reference value from the literature is shown for comparison.

\begin{table}[b]
    \caption{\label{tab:nmq_irt}
    Discrimination ($a$) and difficulty ($b$) metrics estimated from fitting a 2PL IRT model to the ICCs for the 31 dichotomously marked NMQ items (FCI items and Type 1 sub-items only).
    For the 5 items with randomized format (NMQ 03, 23, 25, 34, 36), the analysis was disaggregated by item format.
    Comparison is made against reference values in the literature from \cite{Wang} for original FCI items in MCQ format.
    }
    \begin{ruledtabular}
        \begin{tabular}{lPdPd}
            & \multicolumn{2}{c}{{\textbf{Discrimination ($a$)}}} & \multicolumn{2}{c}{\textbf{Difficulty ($b$)}}\\
            \multicolumn{1}{c}{\textbf{Item \#}} & \multicolumn{1}{c}{\textbf{NMQ}} & \multicolumn{1}{c}{\textbf{Ref. value \cite{Wang}}} & \multicolumn{1}{c}{\textbf{NMQ}} & \multicolumn{1}{c}{\textbf{Ref. value \cite{Wang}}}\\
            \hline
            NMQ 01 & 1.0 + 0.1 & 0.72 & -1.9 + 0.2 & -1.53\\
            NMQ 02 & 1.3 + 0.1 & - & -0.7 + 0.1 & -\\
            NMQ 03 (MCQ) & 1.0 + 0.2 & 0.81 & -0.4 + 0.2 & 0.57\\
            NMQ 03 (FTR) & 1.5 + 0.3 & - & 0.0 + 0.1 & -\\
            NMQ 04 & 2.4 + 0.2 & 2.16 & -0.1 + 0.1 & 1.08\\
            NMQ 05 & 0.7 + 0.1 & - & -2.5 + 0.4 & -\\
            NMQ 06 & 0.6 + 0.1 & - & -1.1 + 0.3 & -\\
            NMQ 07 & 1.6 + 0.2 & 0.59 & -1.5 + 0.1 & -1.7\\
            NMQ 08 & 1.3 + 0.1 & - & -0.8 + 0.1 & -\\
            NMQ 09 & 1.3 + 0.2 & 0.71 & -0.9 + 0.1 & -0.19\\
            NMQ 10 & 1.4 + 0.2 & 0.81 & -1.1 + 0.1 & 0.88\\
            NMQ 11 & 1.8 + 0.2 & 0.93 & -1.2 + 0.1 & -0.72\\
            NMQ 12 & 1.6 + 0.2 & 1.54 & -0.5 + 0.1 & 0.75\\
            NMQ 13 & 1.3 + 0.2 & 0.53 & -1.5 + 0.2 & -0.64\\
            NMQ 14 & 1.1 + 0.1 & 0.66 & -0.6 + 0.1 & -1.32\\
            NMQ 15 & 3.0 + 0.3 & 2.83 & -0.12 + 0.06 & 0.58\\
            NMQ 16 & 1.8 + 0.2 & 1.23 & -1.0 + 0.1 & 0.46\\
            NMQ 17 & 1.1 + 0.1 & - & -1.4 + 0.2 & -\\
            NMQ 18 & 1.4 + 0.2 & 0.59 & -1.7 + 0.1 & -0.81\\
            NMQ 20 & 0.5 + 0.2 & - & -2.1 + 0.5 & -\\
            NMQ 21 & 1.4 + 0.2 & 1.95 & -0.6 + 0.1 & 1.19\\
            NMQ 23 (MCQ) & 1.2 + 0.3 & 1.74 & 0.7 + 0.2 & 1.33\\
            NMQ 23 (FTR) & 0.7 + 0.2 & - & -1.7 + 0.5 & -\\

            NMQ 24 & 1.2 + 0.2 & - & -0.7 + 0.1 & -\\
            NMQ 25 (MCQ) & 1.1 + 0.2 & 0.53 & -1.2 + 0.2 & -0.64\\
            NMQ 25 (FTR) & 0.7 + 0.2 & - & -0.7 + 0.3 & -\\
            NMQ 26 & 1.1 + 0.1 & - & -1.7 + 0.2 & -\\
            NMQ 28 & 1.5 + 0.2 & 2.72 & 0.00 + 0.08 & 0.87\\
            NMQ 29 & 0.5 + 0.1 & - & -0.3 + 0.2 & -\\
            NMQ 30 & 1.4 + 0.2 & 1.44 & -0.44 + 0.09 & 1.39\\
            NMQ 32 & 3.0 + 0.7 & - & -2.2 + 0.1 & -\\
            NMQ 34 (MCQ) & 0.7 + 0.2 & 0.63 & 0.5 + 0.2 & 1.29\\
            NMQ 34 (FTR) & 1.2 + 0.2 & - & 0.3 + 0.2 & -\\
            NMQ 36 (MCQ) & 1.4 + 0.3 & 0.87 & -1.8 + 0.2 & -0.17\\
            NMQ 36 (FTR) & 0.8 + 0.3 & - & -3.0 + 0.9 & -\\
            NMQ 38 & 1.4 + 0.2 & - & 0.50 + 0.08 & -\\
        \end{tabular}
    \end{ruledtabular}
\end{table}


The parameters listed in Table~\ref{tab:nmq_irt} for FCI items in MCQ format have varied agreement with reference values.
The discrimination of FCI items embedded in the NMQ has reasonably close agreement with the reported literature, however, the difficulty of embedded FCI items was generally lower for the NMQ than reported elsewhere.
The average pre-test half-length FCI score for NMQ attempts was 9.42 (out of 14), indicating a strong cohort of students, which may skew the difficulty estimation of NMQ items towards lower values.
However, this still allows relative benchmarking of Type 1 sub-items and FCI items in FTR formats against established FCI items in MCQ format.

\end{document}